\documentclass{article}

\usepackage{amsmath}
\usepackage{amssymb}
\usepackage{theorem}

\usepackage{graphicx}

\if@french \newtheorem{Def}{D\'efinition}
\else \newtheorem{Def}{Definition}
\fi
\if@french \newtheorem{definition}{D\'efinition}
\else 
\fi
\if@french \newtheorem{Theo}{Th\'eor\`eme}
\else \newtheorem{Theo}{Theorem}
\fi
\if@french \newtheorem{theorem}{Th\'eor\`eme}
\else 
\fi
\newtheorem{Prop}[Theo]{Proposition}

\if@french \newtheorem{Lemma}[Theo]{Lemme}
\else \newtheorem{Lemma}[Theo]{Lemma}
\fi
\if@french \newtheorem{Cor}{Corollaire}[Theo]
\else 
\fi

\if@french
        \newenvironment{Proof}[1][Preuve]{\paragraph{{#1}}}%
                {{\hfill\(\Box\)\\}}
\else
        \newenvironment{Proof}[1][Proof]{\paragraph{{#1}}}%
                {{\hfill\(\Box\)\\}}
\fi
\if@french
        \newenvironment{proof}[1][Preuve]{\paragraph{{#1}}}%
                {{\hfill\(\Box\)\\}}
\else
                {{\hfill\(\Box\)\\}}
\fi

\if@french
        \newenvironment{SProof}[1][\'Ebauche de preuve]{\paragraph{{#1}}}%
                {{\hfill\(\Box\)\\}}
\else
                {{\hfill\(\Box\)\\}}
\fi

\if@french
\newenvironment{Example}[1][Exemple]%
        {\paragraph{{#1}}\begin{list}{}{}\item
        }{\end{list}}
\else
        {\paragraph{{#1}}\begin{list}{}{}\item
        }{\end{list}}
\fi

\if@french
        
\else
        
\fi
\if@french
        
\else
        
\fi

\newcommand{\bra}{\langle}
\newcommand{\ket}{\rangle}

\newcommand{\tuple}[1]{\ensuremath{\left\langle{#1}\right\rangle}}
\newcommand{\coll}[1]{\ensuremath{\left\{ {#1}\right\} }}

\newcommand{\sem}[1]{\ensuremath{[\![{#1}]\!]}}

\newcommand{\paren}[1]{\ensuremath{\left( {#1} \right)}}

\newcommand{\set}[2]{\ensuremath{\left\{\left.#1\,\,\vphantom{#2}\right|\,#2\right\}}}
\newcommand{\fall}[1]{{\forall\,{#1},\ }}
\newcommand{\fexist}[1]{{\exists\,{#1}\,{:}\ }}

\newcommand{\mc}[1]{{\mathcal{#1}}}

\newcommand{\mb}[1]{{\bf #1}}

\newcommand{\sas}{\mathbin{\&}}

\newcommand{\feuniq}[1]{{\exists\,!\,{#1}\,{:}\ }}

\begin{document}

\title{Partial Description of Quantum States}
\author{Olivier Brunet\footnote{olibrunet@free.fr}}

\maketitle

\begin{abstract}
	One of the most central and controversial element of quantum mechanics is the use of non zero vectors of a Hilbert space (or, more generally, of one dimension subspaces) for representing the state of a quantum system. In particular, the question whether such a representation is complete has been debated since almost the early days of quantum mechanics. \par	
	In this article, we develop an alternate way to formalize knowledge about the state of quantum systems, based solely on experimentally accessible elements, namely on outcomes of finite measurements. We introduce what we call \emph{partial description} which, given a feasible measurement, indicates some outcomes which are known to be impossible (i.e. known to have a probability equal to $0$ to occur) and hence have to be discarded. Then, we introduce \emph{partial states} (which are partial descriptions providing as much information as possible) and compare this way to describe quantum states to the orthodox one, using vector rays. \par
	Finally, we show that partial states allow to describe quantum states in a strictly more expressive way that the orthodox description does.
\end{abstract}

\section{Introduction} 
\label{sec:philosophical_examinations}

In the standard formulation of quantum mechanics using the Hilbert framework, a quantum system is represented by a complex Hilbert space $\mc H$, in which a state of the system is represented by a non-zero vector $|\varphi\ket$ or, more generally, by its span $\mb C |\varphi\ket $, as a vector represents a state up to phase factor.

\medskip

In particular, let us focus on some elements of the formulation of quantum mechanics using the Hilbert space framework:
\begin{enumerate}
	\item A quantum system $S$ is represented by a complex Hilbert space $\mc H_S$;
	\item A state of $S$ is represented by a unitary vector $|\varphi\ket$ of $\mc H_S$ or, more generally, by a vector ray, i.e. a one-dimensional subspace of $\mc H_S$, corresponding to the span $\mb C |\varphi\ket$ of $|\varphi\ket$;
	\item Information about the state of $S$ is obtained through a processus called \emph{measurement} which, in terms of Hilbert spaces, is represented by a hermitian operator $M$ on $\mc H_S$, and which is postulated to work the following way: if, immediately before a measurement $M$ is applied, the state of a quantum system is $|\varphi\ket$ and if $i$ is an eigenvalue of $M$ and $P_i$ denotes the orthogonal projection on the eigenspace of $M$ associated to $i$, then outcome $i$ will, according to the Born rule, occur with probability~:
	$$ p_i\paren{|\varphi\ket} = \frac {\bra \varphi | P_i | \varphi \ket}{\bra \varphi | \varphi \ket} $$
	In that case, right after the measurement occured, the state of the system will be $ P_i |\varphi\ket $ (up to a normalization factor, which we do not take into account, since we consider that quantum states are actually represented by vector rays and not by vectors themselves).
\end{enumerate}

While probabilities play a central role in quantum mechanics, we will try to avoid them as much as possible in our discussion and, to that respect, we shall retain only one aspect of their meaning: the probability $p_i(\varphi)$ of an outcome $i$ given a state $|\varphi\ket$ indicates whether such an outcome is possible (in which case the probability is different from $0$) or not (the probability equals~$0$).

\medskip

This formulation of quantum mechanics, despite being extremely successfull in its applications, has been problematic from almost its beginning. These difficulties, symbolized by the Einstein-Podolsy-Rosen argument \cite{Einstein35EPR}, come from the use of wave functions, that is the fact that a pure quantum state is represented in the quantum theory by a vector in the corresponding Hilbert space.

Problematic aspects include the question whether such a representation is complete, that is wether such a formalism provides a complete description of the state of a quantum system. If this were the case, then this would mean that in our description of the world, some fundamental assumptions should be given up, such as determinism (the state of a quantum system does not in general specify precisely the outcome of a measurement but only the probability to obtain it) or locality (with the possibility of direct and instantaneous influence between distant objects).

A lot of literature has been devoted to these question with, in particular, the study of the possibility of some ``hidden-variable'' theories, where the description of the state of a quantum system is complemented by some extra elements of information. However, some major results have shown that such hidden variables theories are not possible under reasonable assumptions. The most central ones are the Kochen-Specker theorem \cite{KochenSpecker67}, the Gleason's theorem \cite{Gleason57} and Bell's theorem \cite{Bell64,Bell85Book}.

However, while these theorems show that some ways to solve the problem are impossible, the situation remains rather unsatisfactory, as illustrated by the never-ending wealth of publications on the subject.

\medskip

In the present article, we  attempt to provide some new insights regarding these questions by developing a different approach for representing knowledge about the state of a quantum system. The key word here is ``knowledge": we deliberatly do not try to describe what the state actually is, but instead we attempt to describe what we do know about a state of a quantum system. In other words, our formalism is based on the use of actual elements of information, by which we mean results of actually performable measurement operations. In the case of quantum mechanics, this corresponds to outcomes of finite measures (we must only consider finite ones since one can only manipulate and deal with a finite amount of information). However, in order to be as generic as possible, we will use the formalism of orthomodular lattices which constitute a more general algebraic formalism (one can refer for instance to \cite{Hughes89Book,DallaChiara2001QuantumLogic,Ptak91Book,Stubbe:2007} for more information).


\section{Organization of the Article} 
\label{sec:organization_of_the_article}

We start by defining the entities which will model to finite measures. In the Hilbert framework, this corresponds to the eigenvalues of a hermitian operator with finitely many eigenvalues. More generally, this can be expressed as a maximal collection of mutually orthogonal subspaces of a Hilbert space which can, more generally, be replaced by elements of an orthomodular lattice. Another way to represent such a finite measurement is given by a boolean sub\-algebra of our orthomodular lattice, which is the sub\-algebra spanned by the previous elements.

\bigskip

Next, we define our representation formalism by means of functions which, given a finite measurement, indicate some outcomes which are ruled out by the experimental setup, that is which have a probability of $0$. However, we do not demand that all outcomes with a zero probability should be ruled out and, more importantly, we do not demand that our functions indicate which will be the actual outcome. 
Such functions will be called \emph{partial descriptions}, where the adjective ``partial'' follows from the fact that they only provide partial information about the outcome of a measurement.

\bigskip

We then define an equivalent way to represent partial descriptions, using what we call Sasaki filters, and study some general results concerning the collection of all Sasaki filters of a given orthomodular lattice. In particular, we show that they form a complete atomic lattice.

\bigskip

Finally, we focus on the study of Sasaki filters in the case where our orthomodular lattice is a Hilbert lattice, that is the lattice made of all closed subspaces of a Hilbert space. More precisely, we will focus on atomic Sasaki filters which we call ``partial states".

If the corresponding Hilbert space is of dimension $2$, we show that partial states are such that they provide a definite answer for every possible measurement, which contrasts deeply with the situation in orthodox quantum mechanics where a state is represented by a vector ray which encode the outcome of exactly one possible measurement. In dimension at least $3$, 
we show two results:

\begin{enumerate}
	\item quantum states (i.e. one-dimensional subspaces) can be seen partial states, 
	\item there are partial states which do not correspond to quantum states.
\end{enumerate}

The latter is a very significant result, since it shows that our formalism, which is exclusively based on results of actually performable measurement, 
permits to consider descriptions of a quantum system which are strictly more expressive than those provided by the orthodox formalism.


\section{Finite Measurements} 
\label{sec:finite_measurements}


\subsection{Finite measurements as finite collection of outputs} 
\label{sub:finite_measurements_as_finite_collection_of_outputs}

The basic example of a finite measurement is, in the Hilbert space framework, provided by hermitian operators with finitely many eigenvalues (i.e. finitely many outcomes). With such a hermitian operator, to each possible outcome can be associated an eigenspace, which is a closed subspace of the Hilbert space modelling our quantum system. Moreover, these eigenspaces are pairwise orthogonal and their sum equals the whole Hilbert space.

Our first definition of a finite measurement, based on elements of an orthomodular lattice, directly follows from these considerations. 

\begin{Def}[Finite Measurement]
	A \emph{finite measurement} of an orthomodular lattice $\mc L$ is a finite collection $M=\coll{e_1, \ldots, e_n}$ verifying:
	$$ \hbox{\it 1.}\ \fall i \bot < e_i \qquad \hbox{\it 2.}\ \fall {i \neq j} {e_i} \leq {e_j}^\bot \qquad \hbox{\it 3.}\ e_1 \vee \cdots \vee e_n = \top$$
	Let $\mathop{\rm FinMes}(\mc L)$ denote the collection of finite measurement of $\mc L$.
\end{Def}

Given two finite measurements $M$ and $M'$, we say that $M$ is finer than $M'$ and denote this by $M \leq_{FM} M'$ if~:
$$ \fall {e \in M} \fexist {f \in M'} e \leq f $$
It can be easily shown that, as suggested by the notation, $\leq_{FM}$ defines a partial order on $\mathop{\rm FinMes}(\mc L)$.

\begin{Prop}
	For all $M,M' \in \mathop{\rm FinMes}(\mc L)$, if $M \leq_{FM} M'$, then one has:
	$$ \fall {e \in M} \feuniq {f \in M'} e \leq f $$
\end{Prop}
\begin{Proof}
	Since $M \leq_{FM} M'$, one only needs to prove the uniqueness of $f$. But suppose that there are two distinct elements of $M'$, namely $f_1$ and $f_2$, such that $e \leq f_1$ and $e \leq f_2$. From the definition of a finite measurement, one has $f_1 \leq {f_2}^\bot$ and thus $e \leq f_1 \leq {f_2}^\bot$. Combined with the fact that $e \leq f_2$, this implies that $e = \bot$, which is not possible.
\end{Proof}

This proposition suggests to define for all pairs $(M,M')$ of finite measurement such that $M \leq_{FM} M'$ a function $\pi_{M \leq M'} : M \rightarrow M'$ with maps an element $e$ of $M$ to the unique element $f \in M'$ such that $e \leq f$.


\subsection{Finite measurements as finite boolean sub\-algebras} 
\label{sub:finite_measurements_as_finite_boolean_sub_algebras}

In order to present different approaches to our partial representations of states, we introduce some notations regarding boolean sub\-algebras.

\begin{Def}
	Given an orthomodular lattice $\mc L$, let $\mathop{\rm FBA}(\mc L)$ denote the collection of finite boolean sub\-algebras of $\mc L$.
\end{Def}

There is an obvious relation between finite measurements and finite boolean sub\-algebras, since given a finite measurement $M$, one can define a finite boolean sub\-algebra by $\set{\bigvee E}{E \subseteq M}$. Conversely, given a finite boolean sub\-algebra, the set of its atoms forms a finite measurement.

We also define the following projection operator on finite boolean sub\-algebras, which will play a role similar to that of $\pi_{M \leq M'}$ for elements of $\mathop{\rm FinMes}(\mc L)$:

\begin{Def}
	Given a finite boolean sub\-algebra $\mc B \in \mathop{\rm FBA}(\mc L)$, we define the projection $\pi_{\mc B}$ on $\mc B$ by~:
	$$ \begin{array}{r@{\ }c@{\ }l} \pi_{\mc B}~:\ \mc L & \rightarrow & \mc B \\ x & \mapsto & \bigwedge \set {y \in \mc B}{x \leq y} \end{array} $$
\end{Def}



\section{Partial Descriptions of a Quantum State} 
\label{sec:partial_description_of_a_quantum_state}

As explained in the introduction, our goal is to develop a way to describe (possibly partially) a quantum state by using only elements corresponding to actual knowledge, that is by results of actually realizable experiments.

In quantum mechanics, this corresponds to using eigenspaces of hermitian operators or, equivalently, closed subspaces of the Hilbert space describing our system. More abstractly, this corresponds to elements of the associated Hilbert lattice.

Our partial descriptions will then be defined as follows: given an actually realizable experiment, that is in our context a finite measurement, a partial description provides information about its expected result. Imposing that our description should precisely give the outcome of any measurement seems an unreasonably strong requirement (it is actually impossible, as it follows from results such as the Kochen-Specker theorem or the generalization presented by the author in \cite{Brunet07PLA}. We will develop on this impossibility later in the article). Instead, we demand a weaker condition: that, given a finite measurement, it tells which outcomes may occur or, considering the complement, it provides a list of outcomes which have a probability equal to $0$.

\bigskip

It should be remarked at this point that we only impose that those outcomes which are considered as impossible should have a probability of $0$. That means that some outcomes may be considered as possible even though they have a probability of $0$. In other words, a partial description provides informations about which outcomes will not occur, and not about which outcomes will.

\subsection{Partial Descriptions} 
\label{sub:partial_descriptions}

Following the previous discussion, we define a \emph{partial description} as a function $d$ which associates to each finite measurement $M$ a non-empty subset $d(M) \subseteq M$. Intuitively, $d$ carries the following pieces of information: if measurement $M$ is performed on the quantum system, then the outcome will be an element of~$d(M)$.

\par

Equivalently, in terms of finite boolean sub\-algebras, a partial description can be defined on boolean sub\-algebras as $d(\mc B)=\bigvee d(M_{\mc B})$ where $M_{\mc B}$ is the partial description composed of the atoms of $\mc B$. One then has:
$$d(\mc B) \in \mc B \quad \hbox{and} \quad d(\mc B)\neq \bot$$
Conversely, $d(M_{\mc B})=\set{o \in \mathrm{atoms}(\mc B)}{o \leq d(\mc B)}$, so that one can express partial descriptions equivalently in terms of finite measurements or of finite sub-boolean algebras. It seems that no confusion can be made regarding these two ways to consider partial descriptions, so that in the following, we will denote the two functions the same way.

\bigskip

In the previous section, we have introduced a partial order relation between partial measurements. In order to reflect this relation on partial descriptions, we demand the following requirement:
$$ \fall {M \leq_{FM} N} d(N) = \set{\pi_{M\leq N}(x)}{x \in d(M)} $$
This condition, which may seem to be straightforward, deserves a closer examination. First, given two finite measures $M \leq_{FM} N$, if an element $e \in M$ is considered as a possible outcome for $M$ with regards to a partial description $d$, i.e. if $e \in d(M)$, then $\pi_{M \leq N}(e)$ has to be a possible outcome for $N$~:
$$ M \leq N \Rightarrow \paren{\fall {e \in d(M)} \pi_{M \leq N}(e) \in d(N)} $$
This can be rewritten has~:
$$ \set {\pi_{M \leq N}(e)}{e \in d(M)} \subseteq d(N) $$
However, we require an equality and not just an inclusion. Thus, we also demand that given an element $f$ of $d(N)$, there has to be an element $e \in d(M)$ such that $f = \pi_{M \leq N}(e)$.
This condition is far from being obvious. On the contrary, it seems to us that it reflects an important feature of quantum physics: consider a system which, after performing a measurement $M_1$, has its state lying in an eigenspace $E_1$, and suppose that one performs another measurement $M_2$ compatible with $M_1$ (that is their associated hermitian operators commute). Without loss of generality, one can consider that every eigenspace of $M_2$ is included in one eigenspace of $M_1$. In that case, after performing $M_2$, the quantum system will have its state belong to an eigenspace $E_2$ with $E_2 \subseteq E_1$, in even though one cannot, in general, tell which outcome will be obtained.

This prediction can be made before performing $M_2$ and is a consequence of quantum theory. It is precisely this important property that we try to capture, by saying that for $M \leq N$, if $f \in N$ is a possible outcome, then there has to be a possible $e \in M$ such that $e \leq f$.

\bigskip

In terms of boolean algebras, this condition can be equivalently expressed~as:
$$ \fall {\mc B_1 \subseteq \mc B_2} d(\mc B_1) = \pi_{\mc B_1} \, d(\mc B_2) $$

%
%
%
%


\subsection{Some technical results} 
\label{sub:some_technical_results}

\begin{figure}[ht]
	\centering
		\includegraphics[scale=0.25]{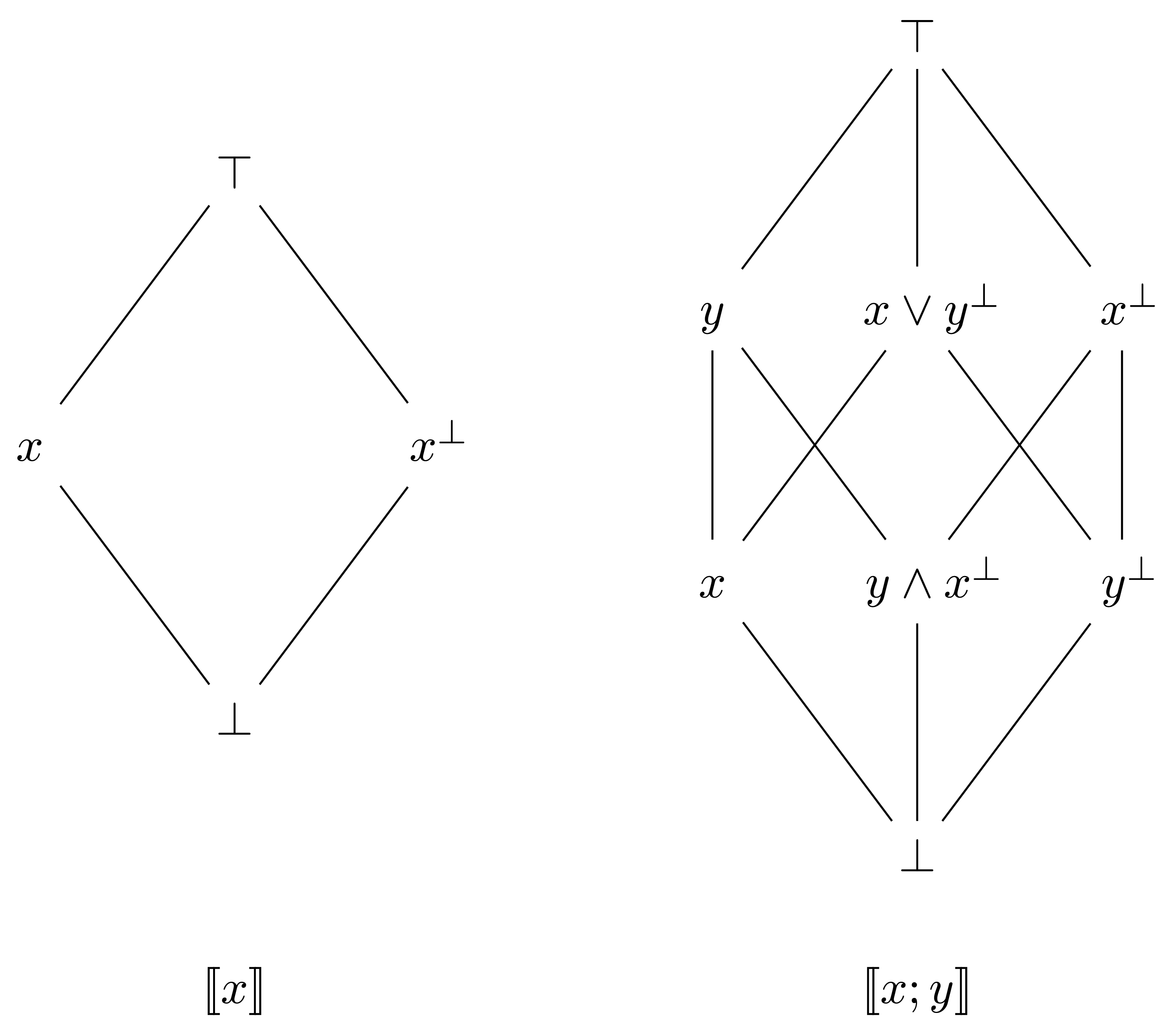}
	\caption{Definition of $\sem x$ and $\sem{x;y}$}
	\label{fig:semnotations}
\end{figure}

In the following, given $x \in \mc L$, let $\sem x$ denote the boolean sub\-algebra generated by $x$, that is~:
$$ \sem x = \coll {\top;x;x^\bot;\bot} $$
Similarly, for $x \leq y$, let $\sem {x;y}$ denote the boolean sub\-algebra generated by $x$ and $y$~:
$$ \sem {x;y} = \coll {\top;y;x \vee y^\bot;x^\bot;x;y \wedge x^\bot;y^\bot;\bot} $$
These sub\-algebras are depicted in figure \ref{fig:semnotations}
\begin{Lemma} \label{lem:sem}
Let $d$ a partial description on $\mc L$ and $\mc B$ a finite boolean sub\-algebra of $\mc L$, and let us define $x = d(\mc B)$. One has $d(\sem x)=x$.
\end{Lemma}
\begin{Proof}
	Since $\sem x \subseteq \mc B$, one has $d(\sem x)=\pi_{\sem x} d(\mc B) = \pi_{\sem x}(x)=x$.
\end{Proof}

The next proposition provides an alternative way to characterize partial descriptions:

\begin{Prop} \label{Prop:AltPartDesc}
	Two following two properties are equivalent:
	\begin{align*}
		& \fall {\mc B_1, \mc B_2 \in \rm{FBA}(\mc L)} \mc B_1 \subseteq \mc B_2 \Rightarrow d(\mc B_1) = \pi_{\mc B_1} d(\mc B_2) & & & (E_1) \\
		& \fall {\mc B_1, \mc B_2 \in \rm{FBA}(\mc L)} d(\mc B_1) \leq \pi_{\mc B_1} d(\mc B_2) & & & (E_2) \\
	\end{align*}
\end{Prop}
\begin{Proof}
	Let us first prove that $(E_1)$ entails $(E_2)$: one has $d(\mc B_1) \leq \pi_{\mc B_1} d(\mc B_2)$ and $d(\mc B_2) \leq \pi_{\mc B_2} d(\mc B_1)$. But if $\mc B_1 \subseteq \mc B_2$, then $\pi_{\mc B_2} d(\mc B_1) = d(\mc B_1)$ so that:
	$$ d(\mc B_2) \leq d(\mc B_1) \leq \pi_{\mc B_1} d(\mc B_2) $$
	By applying $\pi_{\mc B_1}$, one finally has $d(\mc B_1) = \pi_{\mc B_1} d(\mc B_1) = \pi_{\mc B_1} d(\mc B_2)$.
	
	\medskip
	
	Conversely, let $\mc B_1$ and $\mc B_2$ in $\rm{FBA}(\mc L)$ and define $x=d(\mc B_2)$ and $y=\pi_{\mc B_1}(x)$. Of course, $x \leq y$. From lemma \ref{lem:sem}, $d(\sem x)=x$ and since $\sem x \subseteq \sem {x;y}$, it follows that $x=d(\sem x)= \pi_{\sem x} d(\sem{x;y})$. In particular, $d(\sem{x;y}) \leq x$. But since $x$ is an atom of $\sem {x;y}$, this implies that $d(\sem{x;y})=x$. Now, $\sem y \subseteq \sem {x;y}$, so that $d(\sem y)=\pi_{\sem y}d(\sem{x;y}) = \pi_{\sem y}(x)=y$. Finally, one has $y \in \mc B_1$, so that $\sem y \subseteq \mc B_1$ and $y=d(\sem y)=\pi_{\sem y}d(\mc B_1)$ and one can write $ d(\mc B_1) \leq \pi_{\sem y} d(\mc B_1) = y = \pi_{\mc B_1} d(\mc B_2) $.
\end{Proof}


Thus, we have introduced \emph{partial descriptions} as functions which, given a finite measurement, provides information about which outcome to expect. However, its definition as a function $d$ mapping a finite measurement $M$ (resp. a finite boolean sub\-algebra $\mc B$) to a non-empty subset $d(M) \subseteq M$ (resp. a non-$\bot$ element $d(\mc B)$ of $\mc B$) is only one possible representation. In the following, we show that partial descriptions of an orthomodular lattice $\mc L$ can equivalently be represented by some particular subsets of $\mc L$ which we call \emph{Sasaki filters}.

\subsection{Sasaki Filters} 
\label{sub:sasaki_filters}

Sasaki filters are based on the use of the operation called ``Sasaki projection'' which plays an important role in the field of quantum logic \cite{DallaChiara2001QuantumLogic,CausationSmets2001}. It is defined as a function mapping two elements $x$ and $y$ of an orthomodular lattice to the element $(x \vee y^\bot) \wedge y$. Intuitively, it corresponds to an algebraic generalization to orthomodular lattices of the orthogonal projection in Hilbert spaces.

\begin{Def}
	Given an orthomodular lattice $\mc L$, a \emph{Sasaki filter} is a subset $F \subseteq L$ verifying the following two conditions~:
	\begin{align*}
		& \fall {x \in F} \fall {y \in \mc L} x \leq y \Rightarrow y \in F & & \hbox{Upward closure} \\
		& \fall {x, y \in F} x \sas y \in F & & \hbox{\&-Stability}
	\end{align*}
	where we define the Sasaki projection $x \sas y$ as $(x \vee y^\bot) \wedge y$.
	
	Moreover, a Sasaki filter is said to be \emph{proper} it is does not contain the least element $\bot$ of $\mc L$, i.e. if it is not equal to $\mc L$.
\end{Def}

\begin{Prop}
	Given a partial description $d$, its image $F(d)$ is a proper Sasaki filter, where:
	$$ F(d)= \set {d(\mc B)}{\mc B \in \mathop{\rm FBA}(\mc L)} $$
\end{Prop}
\begin{Proof}
	First, let us show that for $x \in F(d)$ and $y \geq x$, one has $y \in F(d)$: if $x \in F(d)$, the following lemma \ref{lem:sem}, $d(\sem x)=x$. Now, $\sem x \subseteq \sem {x;y}$ so that $d(\sem{x;y}) \leq \pi_{\sem x} d(\sem {x;y}) = d(\sem x)$ which implies that $d(\sem {x;y})=x$. Finally, $d(\sem y)=\pi_{\sem y} d(\sem{x;y})=\pi_{\sem y}(x)=y$ so that $y \in F(d)$.
	
	Now, suppose that $x$ and $y$ are both in $F(d)$ and let us show that $x \sas y = (x \vee y^\bot) \wedge y$ is also in $F(d)$: first, it is clear that $x \vee y^\bot \in F(d)$ since $x \in F(d)$ and $x \leq x \vee y^\bot$. Since $y^\bot \leq x \vee y^\bot$, both $y$ and $x \vee y^\bot$ belong to $\mc B = \sem{y^\bot;x \vee y^\bot}$ and one has $d(\mc B) \leq y$ and $d(\mc B) \leq x \vee y^\bot$. This implies that $d(\mc B) \leq x \sas y = y \wedge (x \vee y^\bot)$ and finally that $x \sas y \in F(d)$.
	
	Finally, it is proper, otherwise one would have $d(\mc B)=\bot$ for all $\mc B$.
\end{Proof}

Conversely, we show that given a Sasaki filter, it is possible to define a partial description in a natural way. This is a consequence of the following proposition, proved in \cite{Brunet07PLA}:

\begin{Prop}
	A subset $F$ of an orthomodular lattice $\mc L$ is a Sasaki filter if and only if for every finite boolean sub\-algebra $\mc B \in \mathop{\rm FBA}(\mc L)$, $F \cap \mc B$ is a principal filter of $\mc B$.
\end{Prop}

\begin{Prop}
	Given a proper Sasaki filter $F$ of $\mc L$, the function $d_F$ which maps every finite boolean sub\-algebra $\mc B \in \mathop{\rm FBA}(\mc L)$ to the least element of $F \cap \mc B$ is a partial description on $\mc L$:
	$$ d_F (\mc B) = \min (F \cap \mc B) $$
\end{Prop}
\begin{Proof}
	This follows from the alternate characterization of partial descriptions given in proposition \ref{Prop:AltPartDesc}: let $\mc B_1$ and $\mc B_2$ be in $\mathop{\rm FBA}(\mc L)$. Since $\pi_{\mc B_1} d_F(\mc B_2) \in \mc B_1$, it follows directly that $d_F(\mc B_1) \leq \pi_{\mc B_1} d_F(\mc B_2)$.
\end{Proof}

We then show that one can univocally associate a partial description and a proper Sasaki filter.

\begin{Prop}
	Given a partial description $d$, one has: $$\fall {\mc B \in \rm{FBA}(\mc L)} d(\mc B) = \min (F(d) \cap \mc B)$$
	Conversely, given a proper Sasaki filter $F$, one has: $$F = \set {\min (F \cap \mc B)}{\mc B \in \rm{FBA}(\mc L)}$$
\end{Prop}
\begin{Proof}
	Concerning the first equality, let $d'$ be defined as $d'(\mc B) = \min (F(d) \cap \mc B)$. Since, $d(\mc B) \in F(d) \cap \mc B$, it follows that $d'(\mc B) \leq d(\mc B)$. Conversely, for all $x$ in $F(d) \cap \mc B$, one has $d(\mc B)\leq x$ so that $d(\mc B) \leq d'(\mc B)$. Thus, $d'=d$.
	
	Now, let $F'=\set {\min (F \cap \mc B)}{\mc B \in \rm{FBA}(\mc L)}$. Let $x$ be in $F'$. There exists a $\mc B$ such that $x = \min (F \cap \mc B)$, so that $x \in \min (F \cap \mc B)$ and $x \in F$. This shows that $F' \subseteq F$. Conversely, let $x$ be in $F$. One has $x = \min (F \cap \sem x)$ so that $x \in F'$. This shows that $F \subseteq F'$ and finally that $F=F'$.
\end{Proof}

Thus, we have shown that partial description can be represented by proper Sasaki filters. In the following, we will use this representation to investigate some properties of the set of partial descriptions of an orthomodular lattice.


\subsection{Some properties of the set of Sasaki filters} 
\label{sub:some_properties_of_the_set_of_sasaki_filters}

Let $\mathrm{SF}(\mc L)$ denote the collection of all Sasaki filters of an orthomodular lattice $\mc L$ and $\mathrm{SF}^\star(\mc L)= \mathrm{SF}(\mc L) \setminus \coll {\mc L}$ the collection of its proper Sasaki filters. Since we have shown that Sasaki filters could serve as an equivalent formulation for partial states, it is interesting to study the structure of $\mathrm{SF}(\mc L)$. We give it the structure of a poset by using the reverse inclusion relation:
$$ F_1 \leq F_2 \Leftrightarrow F_2 \subseteq F_1 $$

In the following, we will prove some lattice-theory properties of $\mathrm{SF}(\mc L)$. We invite the reader to refer to classical lattice theory textbooks for more information, such as \cite{Birkhoff67Book,Gratzer78Book,Davey90Book}.

\begin{Prop}
	$\mathrm{SF}(\mc L)$ is a bounded poset.
\end{Prop}
\begin{Proof}
	It is clear that it admits $\coll \top$ as its greatest element, and $\mc L$ as its least element.
\end{Proof}

\begin{Prop}
	$\mathrm{SF}(\mc L)$ is a complete lattice.
\end{Prop}
\begin{Proof}
Given a collection $\coll{F_i}_{i \in \mc I}$ of Sasaki filters, let $F_\vee$ denotes their intersection:
$$F_\vee = \bigcap_{i \in \mc I} F_i$$
and let us prove that $F_\vee$ is the join of $\coll{F_i}_{i \in \mc I}$ in $\mathrm{SF}(\mc L)$:
\begin{enumerate}
	\item $F_\vee$ is in $\mathrm{SF}(\mc L)$, since upward closure and \&-stability are preserved by arbitrary intersection.
	\item For all $i \in \mc I$, one has $F_\vee \subseteq F_i$, that is $F_i \leq F_\vee$.
	\item Let $G$ be in $\mathrm{SF}(\mc L)$, one has:
\begin{eqnarray*}
	\fall {i \in \mc I} F_i \leq G & \Leftrightarrow & \fall {i \in \mc I} G \subseteq F_i \\ & \Leftrightarrow & G \subseteq \bigcap \coll{F_i}_{i \in \mc I} \\  & \Leftrightarrow & F_\vee \leq G
\end{eqnarray*}
\end{enumerate}
Thus, we have shown that any collection $\coll{F_i}_{i \in \mc I}$ of Sasaki filters has a join defined as:
$$ \bigvee \coll{F_i}_{i \in \mc I} = \bigcap \coll{F_i}_{i \in \mc I} $$
Moreover, since $\mathrm{SF}(\mc L)$ has a least element (which is $\mc L$), it is possible to define a meet operation the usual way by:
\begin{eqnarray*}
	\bigwedge \coll{F_i}_{i \in \mc I} & = & \bigvee \set {G \in \mathrm{SF}(\mc L)}{\fall {i \in I} G \leq F_i} \\ & = & \bigcap \set {G \in \mathrm{SF}(\mc L)}{\fall {i \in I} F_i \subseteq G}
\end{eqnarray*}
Thus, $\tuple{\mathrm{SF}(\mc L),\leq,\vee,\wedge}$ is a complete lattice.
\end{Proof}

\begin{Prop} \label{Prop:AtomicSF}
	$\mathrm{SF}(\mc L)$ is atomic.
\end{Prop}
\begin{Proof}
	Given a Sasaki filter $F$, let $\coll{F_i}_{i \in \mc I}$ be a maximal chain of $\mathrm{SF}^\star(\mc L)$ containing $F$, and define:
	$$ F_\infty = \bigcup \coll{F_i}_{i \in \mc I} $$
	It is clear that $F_\infty$ is upward closed. It is also \&-stable: for $x,y \in \bigcup \coll{F_i}_{i \in \mc I}$, there exists an index $i(x,y)$ such that $x,y \in F_{i(x,y)}$ so that $x \sas y \in F_{i(x,y)}$ and finally, $x \sas y \in F_\infty$. Thus, $F_\infty$ is a Sasaki filter.
	
	By maximality of the chain, it is either the least element of $\mathrm{SF}(\mc L)$, i.e. $\mc L$ itself, or an atom of $\mathrm{SF}(\mc L)$, in which case it does not containt $\bot$. But since for all $i$, one has $\bot \not \in F_i$, it follows that $\bot \not \in F_\infty$, so that $F_\infty$ is an atom of $\mathrm{SF}(\mc L)$, and it verifies $F_\infty \leq F$.
\end{Proof}

\begin{Prop}
	The application $x \mapsto x^\uparrow = \set{y \in \mc L}{x \leq y}$ is an order- and join-preserving injection of $\mc L$ in $\mathrm{SF}(\mc L)$.
\end{Prop}
\begin{Proof}
	It is routine to show that for $x \in \mc L$, $x^\uparrow \in \mathrm{SF}(\mc L)$. Moreover, $x \mapsto x^\uparrow$ is clearly injective and order-preserving. It is also join-preserving since:
$$ \begin{array}
	{r@{\ =\ }l}
(x \vee y)^\uparrow & \set z {x \vee y \leq z} = \set z {x \leq z \ \hbox{and}\  y \leq z} \\ & {\set z {x \leq z}} \cap {\set z {y \leq z}} = {\set z {x \leq z}} \vee {\set z {y \leq z}} \\ & x^\uparrow \vee y^\uparrow \end{array} $$
\end{Proof}
It should be remarked that it is, in general, not meet-preserving.

\bigskip

We summarize all theses results in the following theorem.

\begin{Theo} \label{Theo:SasakiProperties}
	Given an orthomodular lattice $\mc L$, the collection $\mathrm{SF}(\mc L)$ of its Sasaki filters ordered by reverse-inclusion is an complete atomic lattice.
	
	Moreover, the application $x \mapsto x^\uparrow = \set{y \in \mc L}{x \leq y}$ is an injection of $\mc L$ in $\mathrm{SF}(\mc L)$ which is order-, meet- but not join-preserving.
\end{Theo}


\subsection{Quantum states and partial descriptions} 
\label{sub:quantum_states_and_partial_descriptions}

In the canonical quantum mechanics formalism, a quantum system is represented by a Hilbert space $\mc H$ and a quantum state is represented by a non-zero vector $|\varphi\ket$ of $\mc H$. More precisely, since a quantum state is given up to a phase factor, a quantum state is represented by the span $\mb C |\varphi\ket$ of $|\varphi\ket$ which is an atom of the associated Hilbert lattice $\mc L_{\mc H}$. Since this formulation of quantum mechanics is supposed to be complete, such an description encodes a maximal amount of information about the state of the system and it is not possible to have a strictly more informative description.

However, we have seen in our study that the collection $\mathrm{SF}^\star(\mc L_{\mc H})$ of all partial descriptions (or equivalently, of all proper Sasaki filters) of $\mc L_{\mc H}$ also possesses extremal elements: its atoms which, from now on, we will call \emph{partial states}. The atomicity of $\mathrm{SF}(\mc L_{\mc H})$ shows that for any partial description, one can find a partial state below it.

In this context, a question which arises naturally is how do these two notions of extremal description, these two notions of state (quantum states and partial states) compare. Are quantum states some sort of partial states? If not, can partial descriptions be seen as approximations of quantum states? In the next section, we will study this question.



\section{Hilbert Lattices} 
\label{sec:hilbert_lattices}

In the following, $\mc H$ will denote a Hilbert space, and $\mc L_{\mc H}$ the associated Hilbert lattice, i.e. the lattice made of the closed subspaces of $\mc H$, partially ordered by inclusion. It is well known that $\mc L_{\mc H}$ is an orthomodular lattice.

\subsection{In Dimension 2} 
\label{sub:in_dimension_2}

We first study the partial states of $\mc L_{\mc H}$ where $\mc H$ is a Hilbert space of dimension $2$. This situation is extremely important in quantum physics and more particularly in the fields of quantum information and quantum computation, since qubits are represented by vectors in $\mb C^2$.

\begin{Prop} \label{Prop:Dim2}
	Let $a$ and $b$ be two one-dimensional subspaces of $\mc H$. Equivalently, $a$ and $b$ are atoms of $\mc L_{\mc H}$. One has either $a=b^\bot$, in which case $a \sas b=\bot$, or $a \sas b = b$.
\end{Prop}
\begin{Proof}
If $a=b^\bot$, then $a \sas b = b \wedge (a \vee b^\bot) = b \wedge b^\bot = \bot$. Otherwise, if $a \neq b^\bot$, then $a \vee b^\bot=\top$ and $a \sas b = b \wedge \top = b$.	
\end{Proof}

\begin{Theo}
	A partial state of $\mc L_{\mc H}$ is a upward-closed subset $F$ of $\mc L_{\mc H}$ which contains exactly one element of each pair of mutually orthogonal atoms of $\mc L_{\mc H}$.
\end{Theo}
\begin{Proof}
	From proposition \ref{Prop:Dim2}, such a subset $F$ is clearly a Sasaki filter: given $a$ and $b$ in $F$, if neither $a=\top$ nor $b=\top$, then $a \sas b=b$ since $a \neq b^\bot$, so that $F$ is \&-stable.
	
	Moreover, it is atomic since it is not possible find a proper Sasaki filter $F'$ such that $F'<F$: if such a $F'$ existed, let $a$ be in $F' \setminus F$. Since $a$ has to be an atom, either $a$ belongs to $F$, which is impossible since $a \in F' \setminus F$, or $a^\bot$ belongs to $F$ which is also impossible since $F'$ is proper.
\end{Proof}

This theorem shows that, contrary to quantum states, partial states in dimension $2$ are such that they carry enough information for telling the outcome of any performable measurement on it: a partial state describing a qubit would encode the result of the measurement of it in any direction.

This situation is extremely different from that of a quantum state which, as we leave probabilites aside, indicates the exact result of only one measurement. Not only quantum states cannot be regarded as partial states, but partial states are infinitely more informative than quantum states.


\subsection{In Dimension 3 and more} 
\label{sub:in_dimension_3_and_more}

However, one can argue that the previous result is not extremely interesting, since a qubit or any system described by a Hilbert space of dimension $2$ is actually part of a bigger system, described by a Hilbert space of higher dimension, that is at least $3$. And $3$ does precisely correspond to the least dimension in which both the Kochen-Specker theorem and the Gleason's theorem hold \cite{KochenSpecker67,Gleason57}.

\subsubsection{Partial states and the Kochen-Specker theorem} 
\label{ssub:atoms_and_kochen_specker}

In \cite{Brunet07PLA}, the author has studied the relation that existed between partial descriptions (refered to as “a priori knowledge”) and the Kochen-Specker theorem which can be stated in terms of proper Sasaki filters as:
\begin{Theo}
	Given a Hilbert space $\mc H$ of dimension at least $3$, there is no proper Sasaki filter $F$ of $\mc L_{\mc H}$ which contains exactly one element of every maximal collection of mutually orthogonal atoms.
\end{Theo}

However, the formulation in terms of proper Sasaki filters also provides a generalization of thie result. Quoting Theorem 7 in \cite{Brunet07PLA}, we have:

\begin{Theo}
	Given a Hilbert space $\mc H$ of dimension at least $3$ and an atom $a$ of $\mc L_{\mc H}$, if proper Sasaki filter $F$ of $\mc L_{\mc H}$ contains $a$, then $F = a^\uparrow$.
\end{Theo}

This result shows that, contrary to what happened in dimension $2$, given an atom $a$ of $\mc L$, the Sasaki filter $a^\uparrow$ is a partial state. Stated another way, to a quantum state $\mb C |x\ket$ can be associated a partial state $(\mb C |x\ket)^\uparrow$ which we call a \emph{principal partial state}.

The next question is now: Are there partial states which are not principal? Are there partial states which do not correspond to quantum states? The next subsection will give a positive answer to this question.


\subsubsection{Non-principal Partial States} 
\label{ssub:there_are_more_states}

Let $\mc H$ be a Hilbert lattice of dimension at least $3$ and let $\coll{|e_i\ket}_{i \in I}$ be an orthomodular basis of $\mc H$. We suppose that $I$ has a particular element, denoted $0$. Moreover, we define a set $\coll{|f_i\ket}_{i \in I}$ of unitary vectors of $\mc H$ by:
$$ |f_0\ket = |e_0\ket \qquad \qquad \fall {i \in I \setminus \coll 0} |f_i\ket = \frac 1 {\sqrt 2} \paren{|e_0\ket + |e_i\ket} $$
It is easy to verify that $\coll{|f_i\ket}_{i \in I}$ is a collection of mutually non-orthogonal vectors, since~:
$$ \fall{i \in I \setminus \coll 0} \bra f_i | f_0 \ket = \frac 1 {\sqrt 2} \qquad \fall{i,j \in I \setminus \coll 0} i\neq j \Rightarrow \bra f_i | f_j \ket = \frac 1 2$$
Now, let us define $G_i = \paren{\mb C |f_i\ket}^\bot = \set {x \in \mc H}{\bra f_i | x \ket = 0}$. One~has:

\begin{Prop} \label{Prop:Counterexample}
	The set $\mc F = \coll \top \cup \bigcup \coll {G_i}_{i \in I}$ is a Sasaki filter of $\mc L_{\mc H}$ and there is no principal maximal Sasaki filter of $\mc L_{\mc H}$ containing $\mc F$.
\end{Prop}
\begin{Proof}
	The fact that $\mc F$ is a Sasaki filter follows from the fact that the elements of $\coll{|f_i\ket}_{i \in I}$ are mutually non-orthogonal. As a consequence, the elements of $\coll {G_i}_{i \in I}$ are mutually incompatible so that if $i \neq j$, then $G_i \sas G_j = G_j$.
	
	Now, suppose that there is an element $|x\ket \in \mc H \setminus \coll 0$ such that $\mc F \subseteq \paren{\mb C|x\ket}^\uparrow$. This means that for all $i$ in $I$, one has $|x\ket \in G_i$ or equivalently that $\fall {i \in I} \bra f_i | x \ket = 0$. In particular, $\bra e_0 | x \ket = \bra f_0 | x \ket=0$ and for $i$ in $\mc I \setminus \coll 0$, one has~:
	$$ \bra e_i | x \ket = \bra \sqrt 2 f_i - e_0 | x \ket  = \sqrt 2 \bra f_i | x \ket - \bra e_0 | x \ket = 0 $$
	This leads to a contradiction, since we assumed that $|x\ket \neq 0$, and as the span of our orthonormal basis $\coll{e_i}_{i \in I}$ is dense in $\mc H$, one has $\fall i \bra e_i | x \ket = 0$ so that $|x\ket=0$.
\end{Proof}

\begin{Theo}
	Given a Hilbert lattice $\mc H$ of dimension at least $3$, there are partial states of $\mc L_{\mc H}$ which are not principal.
\end{Theo}
\begin{Proof}
	This is a direct consequence of propositions \ref{Prop:AtomicSF} and \ref{Prop:Counterexample}.
\end{Proof}




\section{Conclusion and perspectives} 
\label{sec:conclusion_and_perspectives}

In our attempt to develop a formalism for representing knowledge about the state of a quantum system by solely using actual results of measurements, we have introduced \emph{partial descriptions} which, given a performable measurement, provides informations about the expected outcome by discarding some values which are known to be impossible.

In a more general and algebraic approach, we have defined this formalism by using orthomodular lattices (which are a generalization of the collection of closed subspaces of a Hilbert space) and have shown that partial descriptions could be represented by \emph{Sasaki filters}. Then, by studying some structural properties of the collection of all Sasaki filters of a given orthomodular lattice, we have shown in particular that it forms an atomic complete lattice. The atomicity is especially interesting, as it shows the existence of ``maximal'' partial descriptions (maximal in the sense that it is not possible to find a partial description providing strictly more information) which we call \emph{partial states} together with the fact that any partial description can be seen as an approximation of a partial state (or, stated the other way, that any partial description can be refined into a partial state).

Finally, by comparing partial states to quantum states (represented by 1-dimensional subspaces of a Hilbert space or, more algebraically, by atoms of an orthomodular lattice), we have shown that, in dimension $2$ or more (even though the situation is different in dimension $2$ and in greater dimension), the formalism of partial states is strictly more expressive that the  orthodox notion of quantum state.

\bigskip

In this situation, what role could partial descriptions and partial states play in quantum mechanics? Do these mathematical constructions have any meaning or legitimacy?

From an operational point of view, the basic components of partial descriptions are outcomes of feasible measurements. However, partial descriptions provide information about any measurements, even non-compatible ones. Now, since it is not possible to perform non-compatible measurements, this means that, even if partial descriptions do correspond to some ``elements of reality'', they would not be entirely accessible experimentally.

More importantly, should one consider all partial descriptions and all partial states as legitimate, or should one only consider some of them? Following from the study of the structure of the collection of all Sasaki filters of a given orthomodular lattice, it is clear that considering a Hilbert space $\mc H$ and the associated Hilbert lattice $\mc L_{\mc H}$, any partial state of the form $a^\uparrow=\set{x \in \mc L_{\mc H}}{a \leq x}$ with $a$ an atom of $\mc L_{\mc H}$ should be considered, as it constitutes a partial description corresponding to an orthodox quantum state. The question which follows is then, are there other partial descriptions which should be considered as legitimate?



\begin{thebibliography}{DCG01}

\bibitem[Bel64]{Bell64}
John~S. Bell, \emph{On the {E}instein-{P}odolsky-{R}osen paradox}, Physics
  \textbf{1} (1964), 195--200.

\bibitem[Bel87]{Bell85Book}
\bysame, \emph{Speakable and unspeakable in quantum mechanics}, Cambridge
  University Press, 1987.

\bibitem[Bir67]{Birkhoff67Book}
Garrett Birkhoff, \emph{Lattice theory}, 3rd ed., Colloquim Publications,
  American Mathematical Society, 1967.

\bibitem[Bru07]{Brunet07PLA}
Olivier Brunet, \emph{A priori knowledge and the {K}ochen-{S}pecker theorem},
  Physical Letters A \textbf{365} (2007), no.~1-2, 39--43.

\bibitem[DCG01]{DallaChiara2001QuantumLogic}
Maria~Luisa Dalla~Chiara and Roberto Giuntini, \emph{Quantum logic}, Handbook
  of Philosophical Logic (D.~Gabbay and F.~Guenthner, eds.), vol. III, Kluwer,
  2001.

\bibitem[DP90]{Davey90Book}
Brian~A. Davey and Hilary~A. Priestley, \emph{Introduction to lattices and
  order}, Cambridge Mathematical Textbooks, 1990.

\bibitem[EPR35]{Einstein35EPR}
Albert Einstein, Boris Podolsky, and Nathan Rosen, \emph{Can quantum-mechanical
  description of physical reality be considered complete?}, Physical Review
  \textbf{47} (1935), 777--780.

\bibitem[Gle57]{Gleason57}
Andrew Gleason, \emph{Measures on the closed subspaces of a hilbert space},
  Journal of Mathematics and Mechanics \textbf{6} (1957), 885--893.

\bibitem[Gra78]{Gratzer78Book}
George Gratzer, \emph{General lattice theory}, Academic Press, 1978.

\bibitem[Hug89]{Hughes89Book}
R.I.G. Hughes, \emph{The structure and interpretation of quantum mechanics},
  Harvard University Press, 1989.

\bibitem[KS67]{KochenSpecker67}
Simon Kochen and Ernst~P. Specker, \emph{The problem of hidden variables in
  quantum mechanics}, Journal of Mathematics and Mechanics \textbf{17} (1967),
  59--87.

\bibitem[PP91]{Ptak91Book}
Pavel Pt\'ak and Sylvia Pulmannov\'a, \emph{Orthomodular structures as quantum
  logics}, Kluwer, 1991.

\bibitem[Sme01]{CausationSmets2001}
Sonja Smets, \emph{On causation and a counterfactual in quantum logic: the
  sasaki hook}, Logique \& Analyse \textbf{173-175} (2001).

\bibitem[SvS07]{Stubbe:2007}
Isar Stubbe and Bart van Steirteghem, \emph{Propositional systems, {H}ilbert
  lattices and generalized {H}ilbert spaces}, Handbook of Quantum Logic and
  Quantum Structures: Quantum Structures (K.~Engesser, D.~M. Gabbay, and
  D.~Lehmann, eds.), Elsevier, 2007, pp.~477--524.

\end{thebibliography}

\providecommand{\bysame}{\leavevmode\hbox to3em{\hrulefill}\thinspace}
\providecommand{\MR}{\relax\ifhmode\unskip\space\fi MR }
\providecommand{\MRhref}[2]{%
  \href{http://www.ams.org/mathscinet-getitem?mr=#1}{#2}
}
\providecommand{\href}[2]{#2}

\end{document}